\documentclass[12pt]{article}
\usepackage{a4,graphicx}
\usepackage[utf8]{inputenc}
\usepackage{amsmath,amsthm,amsfonts,amssymb,amsopn,amscd}     

\newtheorem{theorem}{Theorem}[section]

\newtheorem{mylemma}[theorem]{Lemma}
\newtheorem{mypropo}[theorem]{Proposition}
\newtheorem{mycoro}[theorem]{Corollary}

\theoremstyle{definition}

\def\citeb{\cite}

\def\dim{{\mathrm{dim}}}

\def\Hom{{\mathrm{Hom}}}
\def\id{{\mathrm{id}}}

\def\eins{{\mathbf{1}}}

\newcommand{\uphook}{\rotatebox[origin=c]{20}{$\hookrightarrow$}\mkern2mu}
\newcommand{\downhook}{\rotatebox[origin=c]{-20}{$\hookrightarrow$}\mkern2mu}

\def\RR{{\mathbb R}}
\def\CC{{\mathbb C}}

\def\A{{\mathcal A}}
\def\B{{\mathcal B}}
\def\C{{\mathcal C}} 
\def\DHR{{\mathrm{DHR}}}

\def\F{{\mathcal F}}

\def\sig{\sigma}
\def\eps{\varepsilon} 
\def\LRA{\Leftrightarrow}

\def\can{_{\mathrm{can}}}
\def\opp{^{\mathrm{opp}}}

\def\ol{\overline}

\newcommand{\bea}{\begin{eqnarray}} 
\newcommand{\eea}{\end{eqnarray}}

\newcommand{\eref}[1]{Eq.~(\ref{#1})}

\newcommand{\cref}[1]{Cor.~\ref{#1}}
\newcommand{\lref}[1]{Lemma~\ref{#1}}

\newcommand{\sref}[1]{Sect.~\ref{#1}}

\begin{document}

\title{Boundaries in relativistic quantum field theory
\footnote{To appear in the proceedings of the XVIII International
  Congress on Mathematical Physics, Santiago de Chile, July 2015}}

\author{Karl-Henning Rehren \\ \small
Institut f\"ur Theoretische Physik, Universit\"at G\"ottingen,
\\[-1mm] \small Friedrich-Hund-Platz 1, D-37077 G\"ottingen, Germany
\\[-1mm] \small E-mail: {\tt rehren@theorie.physik.uni-goettingen.de}}

\maketitle

\begin{abstract}
Boundary conditions in relativistic QFT can be classified by deep
results in the theory of braided or modular tensor categories.
\end{abstract}

\bigskip


\section{The physics problem}
\setcounter{equation}{0}

We study the behaviour of relativistic quantum field theory (QFT) in
the presence of a spacelike boundary (= a hypersurface with spacelike
normal). The local fields on both sides of the boundary are defined on 
the same Hilbert space. The boundary is assumed to be ``transparent'' 
for certain quantum fields, including the stress-energy tensor (SET). 
This means that the fields on both sides share the same SET. 
(In two-dimensional conformal QFT, this property is equivalent to 
conservation of energy and momentum at the boundary \citeb{BKLR}.)  

Because the SET provides the generators for translations, it
can be used to extend the fields, a priori supported only on
one side of the boundary, to all of Minkowski spacetime. One 
therefore has two QFTs on the same Hilbert space, called
$\B^L$ and $\B^R$, that share a common subtheory $\A$, and a common
covariance.   

The principle of causality only requires that the original local
observables commute when they are spacelike separated. In two
dimensions, this implies that the extended ``left'' fields commute with the
``right'' fields whenever the former are localized in the spacelike
left of the latter (``one-sided locality'') -- but not vice versa. In
four dimensions, using Lorentz covariance, they must be relatively local. 

Because the interesting new feature is one-sided locality, we restrict
to two dimensions. The question to be addressed is therefore: how can
a given subtheory $\A$ be embedded into a pair of local extensions $\B^L$
and $\B^R$ such that $\B^L$ is left-local w.r.t.\ $\B^R$. Yet another
way to look at the problem is to consider the covariant QFT $\C$
generated by $\B^L$ and $\B^R$, which is in general non-local, but
relatively local w.r.t.\ $\A$. Then $\B^L$ and $\B^R$ are
intermediate extensions, such that the diagram commutes:
\bea\label{square}
\begin{array}{ccccc}
&&\B^L&&\\[-2.4mm]
&\uphook&&\downhook&
\\[-2.4mm] \A &&&& \C\\[-2.4mm]
&\downhook&&\uphook&
\\[-2.4mm] &&\B^R&&
\end{array},
\eea
and the embedded $\B^L$ is left-local w.r.t.\ the embedded $\B^R$, 
and both generate $\C$.  

When $\A\hookrightarrow\B^L$ and $\A\hookrightarrow \B^R$ are given, 
a ``boundary condition'' between $\B^L$ and $\B^R$ is a 
realization of this diagram. 
It is called irreducible if $\A'\cap\C=\CC\cdot\eins$. 
If $\A\hookrightarrow\B^L$ and $\A\hookrightarrow\B^R$ are isomorphic, the trivial solution is to identify $\B^L=\B^R$. 
Nontrivial solutions can be studied in terms of representation theory of $\A$. 

Because boundary conditions must be algebraically consistent with commutation relations required by causality, they cannot be 
{\em imposed} as in classical field theory. Instead, 
a highly nontrivial classification emerges in the case of
completely rational conformal QFT. 

\section{The mathematical setup}

A QFT is described in terms of its (local and covariant) net of local algebras
$$O\mapsto \A(O).$$
Here $O$ are bounded open spacetime regions (it is sufficient to
consider ``double-cones'' which are the intersections of a forward and a
backward lightcone), and $\A(O)$ is the von Neumann algebra
of observables accessible in the region $O$. Thus, we are
looking for boundary conditions as covariant simultaneous realizations of 
\bea\label{localsquare}
\begin{array}{ccccc}
&&\B^L(O)&&\\[-2.4mm]
&\uphook&&\downhook&
\\[-2.4mm] \A(O) &&&& \C(O)\\[-2.4mm]
&\downhook&&\uphook&
\\[-2.4mm] &&\B^R(O)&&
\end{array}
\eea
for all double-cones $O\subset \RR^2$, such that $B^L(O_1)$ commutes 
with $B^R(O_2)$ whenever $O_1$ is in the left component of the
causal complement of $O_2$. 

The vacuum representation of $\C$ is a reducible positive-energy
representation of $\A$, containing the vacuum representations of 
$\B^L$ and $\B^R$. Thus, boundary conditions are an issue of 
positive-energy representations. 

Positive-energy representations are efficiently described by the
DHR theory \citeb{DHR}, which realizes them as {\em localized
  endomorphisms $\rho$ of the quasilocal algebra $\A$}. 
These are the objects of a C* tensor category equipped with a unitary 
braiding, called $\DHR(\A)$. Intertwiners $t\in\Hom(\rho,\sig)$ are 
elements of $\A$ satisfying $t\rho(a)=\sig(a)t$ for all $a\in\A$. The monoidal 
product of endomorphisms is the composition $\rho\sig$, which
canonically induces the monoidal product of intertwiners. The braiding is
a collection of intertwiners $\eps_{\rho,\sig}\in\Hom(\rho\sig,\sig\rho)$ 
defining a natural isomorphism, as functors 
$\DHR(\A)\times\DHR(\A)\to\DHR(\A)$, between the monoidal product and its 
reversed. The braiding was originally designed to describe the statistics of 
scattering states in massive QFT \citeb{DHR}. In two-dimensional conformal 
QFT, its relation with the exchange properties of conformal blocks was 
established in \citeb{FRS}. 

Its presence is due to the fact that by locality of $\A$, DHR endomorphisms 
commute whenever they are localized at spacelike distance. Thus, putting 
\bea\label{trivialization}
\eps_{\rho,\sig}\stackrel\cdot=\eins
\eea
whenever $\rho$ is localized in the spacelike right of $\sigma$ 
(in two dimensions), consistently defines $\eps_{\rho,\sig}$ in 
the general case by demanding naturality \citeb{DHR}. (The choice of 
``right'' is just a matter of convention, cf.\ \citeb{FRS}; the choice 
of ``left'' would define the opposite braiding 
$\eps_{\rho,\sig}\opp=(\eps_{\sig,\rho})^*$.) 

In two dimensions, double-cones are of the form $O=I\times J$ in lightlike 
coordinates $t\pm x$. We specify $\A$ in \eref{localsquare} to be a conformal QFT with local algebras 
$$\A(O)=\A^+(I) \otimes 
\A^-(J),$$ i.e., the common subtheory consists only of local chiral observables.
Then one has 
$$\DHR(\A)=\DHR(\A^+)\boxtimes\DHR(\A^-)\opp,$$
i.e., the objects of $\DHR(\A)$ are (equivalent to) direct sums of tensor 
products of DHR endomorphisms of $\A^+$ and $\A^-$, equipped 
with the tensor product $\eps\otimes\eps\opp$ of chiral braidings (defined analogously by replacing ``right'' with ''lightlike future''). 
The opposite braiding $\eps\opp$ of $\A^-$ arises because $\rho^+\otimes\rho^-$ is localized in 
the spacelike right of $\sig^+\otimes \sig^-$ iff $\rho^+$ is localized in the 
future of $\sig^+$ and $\rho^-$ in the past 
of $\sig^-$. 

\section{Extensions, Q-systems, and boundary conditions}
\label{s:ext}

Relatively local covariant extensions $\A\hookrightarrow\B$ of a local
quantum field theory $\A$ were classified in \citeb{LR95} in terms of
$\DHR(\A)$. They are in one-to-one correspondence (up to equivalence)
with Q-systems = triples $(\Theta,W,X)$ where $\Theta$ is a DHR
endomorphism equivalent 
to the vacuum representation of $\B$ regarded as a representation of
$\A$, and $W\in\Hom(\id,\Theta)$ and $X\in \Hom(\Theta,\Theta^2)$ are
a pair of intertwiners satisfying the relations of a Frobenius algebra
in the C* tensor category $\DHR(\A)$. $\A\hookrightarrow\B$ is
irreducible iff $\Theta$ contains $\id$ (the vacuum representation of
$\A$) with multiplicity one. $\B$ is local iff $\eps_{\Theta,\Theta}X=X$ 
(i.e., the Q-system is commutative). The algebraic relations of $\B$ as 
well as its local subalgebras $\B(O)$ are encoded in the Q-system. 

For the problem at hand, one looks for Q-systems for
$\A\hookrightarrow \C$ which (a) contain ``intermediate'' Q-systems
(see \citeb[Ch.\ 4.4]{Tcats}) for $\A\hookrightarrow \B^Y\hookrightarrow \C$ ($Y=L,R$), and
(b) whose algebraic relations ensure left locality of $\B^L$ w.r.t.\ $\B^R$. 
Our first result is 
\begin{mypropo} \label{universalconstruction} \emph{\bf (\citeb[Prop.\ 5.1]{BKLR})} 
Given two irreducible Q-systems for $\A\hookrightarrow\B^L$,
$\A\hookrightarrow\B^R$, there is a ``universal construction'' of a Q-system
for $\A\hookrightarrow\C$, such that the central decomposition of\/ $\C$
gives all inequivalent irreducible boundary conditions. 
\end{mypropo} 

The universal construction is the {\em braided product of extensions} 
$\C=\B^L\times^-\B^R$. This is the extension defined by the 
{\em braided product of Q-systems} (cf.\ \citeb[Sec.\ 3.2]{FFRS}) 
$Q^L\times^-Q^R \equiv (\Theta,W,X)$, which is defined by 
$\Theta=\Theta^L\Theta^R$, $W=W^L\times W^R$, and  
$$X:=(1_{\Theta^L}\times\eps_{\Theta^L,\Theta^R}\opp\times
1_{\Theta^R})\cdot (X^L\times X^R).$$ 
The requirement of left locality dictates the choice of the braided product 
$\times^-$ involving the {\em opposite} braiding: the algebraic relations 
of this product Q-system include the commutation relations 
\bea\label{commutationrelations}
\psi_\sig^R\psi_\rho^L = \eps_{\rho,\sig}\opp\cdot
\psi_\rho^L\psi_\sig^R,
\eea
and the opposite braiding is trivial, by \eref{trivialization},
precisely when $\rho$ is localized in the spacelike left of
$\sig$. Here, $\psi^Y_\rho$ ($Y=L,R$) are charged generators of $\B^Y$
carrying irreducible charge $\rho\prec\Theta^Y$ in $\DHR(\A)$, 
and inheriting the localization of $\rho$. 

The universality of this construction follows from the fact that 
the commutation relations \eref{commutationrelations} are the only 
independent algebraic relations defining $\C$, besides the (given)  
algebraic relations of the generators $\psi^Y$ within $\B^Y$ ($Y=L,R$);
whereas left locality -- the only apriori condition for the intermediate 
embeddings of $\B^L$ and $\B^R$ -- requires \eref{commutationrelations} 
whenever $\rho$ is localized to the left of $\sig$, and naturality of 
the braiding implies \eref{commutationrelations} also in the general case.

\smallskip

We write $\iota^Y:\A\to\B^Y$ ($Y=L,R$) and $\iota:\A\to\C$ for the 
inclusion homomorphisms, and $\overline\iota^Y:\B^Y\to\A$, $\ol\iota:\C\to\A$ 
for their conjugates, such that $\ol\iota^Y\iota^Y=\Theta^Y$, and 
$\ol\iota\,\iota=\Theta=\Theta^L\Theta^R$ \citeb{LR95}. The irreducible 
decomposition of $\A\hookrightarrow\C$ is given by the minimal projections 
in the relative commutant $\A'\cap \C\equiv\Hom(\iota,\iota)$. But one has
\begin{mylemma}\label{localtimeslocal} \emph{\bf (\citeb[Prop.~4.33]{Tcats})} 
If \/ $\C$ is the braided product of two \underline{local} extensions 
$\A\hookrightarrow \B^Y$ ($Y=L,R$), then $\A'\cap\C = \C'\cap \C$. 
In particular, the irreducible boundary conditions are classified by 
the minimal central projections of $\C$. Moreover, the centre 
$\C'\cap\C=\Hom(\iota,\iota)$ and $\Hom(\iota^L\ol\iota^R,\iota^L\ol\iota^R)$ 
are isomorphic as algebras.
\end{mylemma}
For the last statement, notice that the two spaces have the same 
images under the embeddings by monoidal units into 
$\Hom(\Theta^L\Theta^R,\Theta^L\Theta^R)$ (using \citeb[Lemma~3.16]{Tcats}).
Thus, the minimal projections in $\C'\cap\C$ (= boundary conditions) 
correspond to the irreducible subhomomorphisms of $\iota^L\ol\iota^R:\B^R\to\B^L$.
Because subhomomorphisms of $\iota^L\rho\,\ol\iota^R:\B^R\to\B^L$, 
where $\rho\in\DHR(\A)$, are $Q^L$-$Q^R$-bimodules \citeb[Ch.~3.6]{Tcats}, 
we conclude that boundary conditions are special $Q^L$-$Q^R$-bimodules. 

\smallskip

The centre of $\C$ is spanned by operators 
$B_\rho\equiv\psi^{L*}_\rho\psi^R_\rho$ such that $\rho\prec\Theta^L$ 
and $\rho\prec\Theta^R$ (suppressing possible multiplicities). 
Every minimal projection assigns a numerical value to $B_\rho$.

To determine the minimal central projections of $\C$, one has to compute and
diagonalize the {\em algebra} of the generators $B_\rho$. One has
\begin{mylemma}\label{astproduct} \emph{\bf (\citeb[Ch.~4.12]{Tcats})} 
Let $Q^Y=(\Theta^Y,W^Y,X^Y)$ ($Y=L,R$) be two commutative
Q-systems. Define the commutative $\ast$-product on $\Hom(\Theta^R,\Theta^L)$
$$T_\rho\ast T_\sig = X^{L*}\cdot(T_\rho\times
  T_\sig)\cdot X^R.$$
There is a linear bijection $\chi:\Hom(\Theta^R,\Theta^L)\to\C'\cap\C$,
taking (appropriate multiples of) matrix units $T_\rho\in
\Hom(\rho,\Theta^L)\cdot\Hom(\Theta^R,\rho)\subset\Hom(\Theta^R,\Theta^L)$ to
$B_\rho$, such that  
$$\chi(T_1)\chi(T_2)=\chi(T_1 \ast T_2).$$  
\end{mylemma}
Thus, diagonalizing the $\ast$-product, diagonalizes
$\C'\cap\C$. If $I_m$ are the minimal projections w.r.t.\ $\ast$, then
$E_m:=\chi(I_m)$ are the minimal projections in $\C'\cap\C$. Expanding 
$$T_\rho=\sum\nolimits_m c_{\rho,m}I_m \qquad \LRA \qquad B_\rho=\sum\nolimits_m
c_{\rho,m}E_m,$$
it follows that $B_\rho\equiv\psi^{L*}_\rho\psi^R_\rho$ take the
numerical values $c_{\rho,m}\in\CC$ in the subrepresentation $\pi_m$ 
of the universal construction given by the
range of $E_m$. These sesquilinear relations among the charged fields
$\psi^L$ and $\psi^R$ are the desired boundary conditions.

\section{Classification of irreducible boundary conditions: modular case}
\label{s:modclass}

The diagonalization of the $\ast$-product in \lref{astproduct} can be
achieved in some special cases. The most remarkable instance is derived 
under the following assumptions. 

\begin{enumerate} \itemsep1mm
\item[(i)] The chiral subtheories $\A^+$ and $\A^-$ have isomorphic DHR 
categories with finitely many irreducible objects of finite dimension.
\item[(ii)] The braiding of $\DHR(\A^\pm)$ is non-degenerate.   
\item[(iii)] Both $\B^L$ and $\B^R$ are maximal local extensions of 
$\A=\A^+\otimes \A^-$.  
\end{enumerate}

The dimension in (i) is the statistical dimension \citeb{DHR,FRS}. 
By (i), there is a canonical commutative Q-system $R\can$ in
$\DHR(\A^+)\boxtimes\DHR(\A^-)\opp$ with
$$\Theta\can\simeq\bigoplus\nolimits_\rho\rho\otimes\overline\rho$$
of dimension $\mu:=\dim(\Theta\can)=\sum_\rho\dim(\rho)^2$,  
where the sums run over the equivalence classes of irreducible objects
(sectors) of $\DHR(\A)$, cf.\ \citeb{LR95}. 

(ii) is an automatic consequence if the chiral theories $\A^\pm$ are
completely rational \citeb{KLM}. By (i) and (ii), $\DHR(\A^\pm)$ is
a modular tensor category \citeb{FRS,KLM}. This implies that the trace
w.r.t.\ $\rho$ of the monodromy
$\eps_{\sig,\rho}\eps_{\rho,\sig}\in\Hom(\rho\sig,\rho\sig)$, summed 
over all sectors $\rho$ of $\DHR(\A^\pm)$, is the projection onto
$\id\prec\sigma$ in $\Hom(\sig,\sig)$, for every $\sig\in\DHR(\A^\pm)$. 
This feature is known as the ``killing ring'' trick (cf.\ \citeb{BEK}).
Still by (i) and (ii), the maximal irreducible commutative Q-systems
are of the form  
$$Q=Z[q]:=C^+[(q\otimes\eins)\times^+ R\can],$$
called the {\em full centre} of $q$ \citeb{KR}. Here, $q$ is an irreducible chiral 
Q-system in $\DHR(\A^+)$, and $\eins$ the trivial Q-system in $\DHR(\A^-)$. 
The (left and right) centres $C^\pm[\cdot]$ of a Q-system \citeb{FFRS} 
are maximal commutative intermediate Q-systems. The full centre is a Morita 
invariant of Q-systems in modular categories \citeb{FFRS,KR}. 
The first characterization of maximal commutative Q-systems 
in terms of the so-called
$\alpha$-induction construction \citeb{R00} was later recognized to
coincide with the full centre \citeb{BKL}. This
also means that their local algebras $\B(O)$ are relative commutants of
nested wedge algebras of the nonlocal braided product extensions $(q\otimes
\eins)\times^+ R\can$ \citeb[Cor.~4.18]{BKLR}.
 
By (i)--(iii), $\B^L$ and $\B^R$ are full centre extensions of
$\A^+\otimes\A^-$, induced by chiral Q-systems $q^L$ and $q^R$. 
To classify the boundary conditions, we first use
\begin{mypropo}\label{bimodules} \emph{\bf (\citeb{FFRS,Tcats})}
Let $q^i=(\theta^i,w^i,x^i)$ be Q-systems in a
braided tensor category. Then $q^i$-$q^j$-bimodules $m$ specify
intertwiners $D_m\in\Hom(\theta^j,\theta^i)$ such that the map $m\mapsto D_m$ 
is invariant under equivalences of bimodules and respects direct sums,
conjugation and bimodule products, normalized as $w^{i*}\cdot D_m\cdot
w^j=\dim(m)$. 
\end{mypropo}
The precise formulation of the statements can be found in \citeb{FFRS,Tcats}. 

Next, the full centre construction ``lifts'' $q^i$-$q^j$-bimodules $m$ to
$Z[q^i]$-$Z[q^j]$-bi\-modules $Z[m]$. We write 
$Z[q^i]=(\Theta^i,W^i,X^i)$. Then one has
\begin{theorem} \emph{\bf (\citeb{FFRS,KR,Tcats})}
For irreducible Q-systems $q^i$ and $q^j$ in a modular C* category,
let $m$ run over the irreducible equivalence classes of 
$q^i$-$q^j$-bimodules. Then 
$$I_m:= \frac{\dim(m)}{\mu\cdot\dim(\theta^i)\dim(\theta^j)}\cdot D_{Z[m]}$$ 
diagonalize the $\ast$-product in $\Hom(\Theta^j,\Theta^i)$ (cf.\
\lref{astproduct}).   
\end{theorem}
Combining this result with \lref{localtimeslocal} and \lref{astproduct}, 
we conclude: 
\begin{mycoro}
The minimal central projections of the universal construction for two full centre extensions are given by $E_m=\chi(I_m)$, where $m$ runs
over the irreducible equivalence classes of chiral $q^L$-$q^R$-bimodules. 
In other words: the boundary conditions, viewed as 
$Z[q^L]$-$Z[q^R]$-bimodules, are induced by the chiral bimodules.
 
\end{mycoro}

In the case $q^L=q^R=\eins$ (or Morita equivalent), hence $Z[q^L]=Z[q^R]=R\can$,
the chiral bimodules are given by $\sig\in\DHR(\A^\pm)$, and one finds
the numerical values  
\bea
\pi_\sig\Big[\psi^{L*}_{\rho\otimes\overline\rho}\psi^R_{\rho\otimes\overline\rho}\Big]=
\frac{\mu^{1/2}}{\dim(\sig)\dim(\rho)}\cdot S_{\sig,\rho} =
\frac{S_{\id,\id}S_{\sig,\rho}}{S_{\id,\sig}S_{\id,\rho}},
\eea
given by the entries of the Verlinde matrix $S$ (if
$\psi_{\rho\otimes\overline\rho}$ are normalized as isometries). 

\section{Other cases}
\label{s:other}

In the general case, one cannot benefit from properties
of modular categories. One may directly read off the $\ast$-product
$T_\rho\ast T_\sig$ from the coefficients of the intertwiners $X^Y$
($Y=L,R$), but a general formula for its minimal projections is not known. 

Special cases can be treated with group theory. If $\DHR(\A)$ is a
symmetric category (e.g., in four dimensions), then it is equivalent
to the dual of a (finite) group, and there is an extension $\F$
with a faithful action of $G$ such that $\A=\F^G$, see \citeb{DR}. The
$\F$-$\F$-boundary conditions are classified by the elements $g\in G$,
and 
\bea
\pi_g\big[\psi^{L*}_{\rho,i}\psi^{R}_{\rho,j}\big] = u^\rho(g)_{ij} \quad
\Rightarrow \quad \pi_g\big[\psi^{R}_{\rho,j}\big] =
\sum\nolimits_i\pi_g\big[\psi^L_{\rho,i}\big]\cdot u^\rho(g)_{ij},
\eea
i.e., the boundary conditions are gauge transformations \citeb{BKLR}. 

If, in the two-dimensional case, $\DHR(\A^+)\simeq\DHR(\A^-)$ contains
a symmetric subcategory (i.e., if the chiral observables admit an
orbifold construction $\A^\pm=(\B^\pm)^G$ with the faithful action of
a finite group), and if $\B^L\simeq\B^R$ are given by the canonical
Q-system of this subcategory, then one finds the $\ast$-product
$$\dim(\rho)T_{\rho\otimes\overline\rho}\ast \dim(\sig)T_{\sig\otimes\overline\sig} =
\sum\nolimits_\tau N_{\rho,\sig}^{\tau} \cdot\dim(\tau)
T_{\tau\otimes\overline\tau},$$
where $N_{\rho,\sig}^{\tau}$ are the fusion rules of the dual of
$G$. This is diagonalized by 
$$\dim(\rho)T_{\rho\otimes\overline\rho} = \sum\nolimits_C \langle \rho\vert
C\rangle \cdot I_C,$$
where the sume runs over the conjugacy classes of $G$ and $\langle \rho\vert
C\rangle$ is the character of $C$ in the representation
$\rho$. Namely, $\langle \rho\vert C\rangle \langle \sig\vert
C\rangle = \sum\nolimits_\tau N_{\rho,\sig}^{\tau}\langle \tau\vert
C\rangle$. Thus, we have 
\bea
\pi_C\Big[\psi^{L*}_{\rho\otimes\overline\rho}\psi^R_{\rho\otimes\overline\rho}\Big]
= \frac{\langle \rho\vert C\rangle}{\dim(\rho)}.
\eea

\section{Juxtaposition of boundaries}
\label{s:jux}

For the juxtaposition of two boundaries between three QFTs $\B^1$, 
$\B^2$, $\B^3$, one expects a composition of boundary conditions. Two
options may be considered.

Since boundary conditions are (special) $Q^i$-$Q^j$-bimodules (\sref{s:ext}), the first 
option is the bimodule tensor product, or equivalently the composition 
of homomorphisms $\alpha^{12}\prec \iota^1\ol\iota^2:\B^2\to\B^1$ and 
$\alpha^{23}\prec \iota^2\ol\iota^3:\B^3\to\B^2$. In general, this does not close 
among boundary conditions, because $\beta^{13}\prec\alpha^{12}\alpha^{23}$ 
are a priori only subhomomorphisms of $\iota^1\Theta^2\,\ol\iota^3$. 
Instead, the bimodule tensor product closes among {\em defects}, 
cf.\ \citeb{BDH,BKLR}, that relax the condition
that $\C$ in \eref{square} is generated by $\B^L$ and $\B^R$. 

Another option is to define the composition of boundary conditions as
the composition of intertwiners $I_m\in\Hom(\Theta^j,\Theta^i)$. This
closes among boundary conditions, but fails to give rise to a tensor
category. One may expect only a fusion ring, like the
{\em product of conjugacy classes} in the second example of \sref{s:other}.

In the case of full centres of a modular CFT (\sref{s:modclass}), both options coincide.

\medskip

{\bf Acknowledgement.} This report is based on joint work with M. Bischoff, Y. Kawahigashi and R. Longo.


\end{document}